\documentclass[aps,prl,reprint,preprintnumbers,nofootinbib,superscriptaddress]{revtex4-2}

\usepackage{amsmath,amssymb}
\usepackage{bm}
\usepackage{booktabs}
\usepackage{graphicx}
\usepackage[colorlinks=true,allcolors=blue]{hyperref}

\newcommand{\PT}{P_T}
\newcommand{\PbarT}{\bar{P}_T}

\begin{document}
\preprint{LA-UR-26-28792}

\title{The Fate of Collisional Neutrino Flavor Conversion: Fixed Points, Period Doubling, and Chaos}

\author{Jiabao Liu}
\email{kamaelliu@asagi.waseda.jp}
\affiliation{Department of Physics and Applied Physics, School of Advanced
Science \& Engineering, Waseda University, Tokyo 169-8555, Japan}

\author{Lucas Johns}
\affiliation{Theoretical Division, Los Alamos National Laboratory, Los Alamos, NM 87545, USA}

\author{Hiroki Nagakura}
\affiliation{Theoretical Astrophysics, Department of Earth \& Space Science, Graduate School of Science, The University of Osaka, 1-1 Machikaneyama, Toyonaka, Osaka 560-0043, Japan}
\affiliation{Division of Science, National Astronomical Observatory of Japan, 2-21-1 Osawa, Mitaka, Tokyo 181-8588, Japan}

\begin{abstract}
Core-collapse supernovae and neutron star mergers are among the most energetic astrophysical events and promising sources of multimessenger signals. Their dynamics and nucleosynthesis depend sensitively on neutrino transport, yet the impact of neutrino flavor evolution in these dense environments remains poorly understood. Predicting the outcome of collisional flavor instability (CFI) remains a challenge, despite progress in local-relaxation methods for fast flavor conversion. Here we construct homogeneous collective-precession states with stationary flavor populations in two-flavor quantum kinetics with collisional relaxation. The method applies to arbitrary energy spectra and energy-dependent collision rates. Our numerical examples exhibit persistent oscillations and a period-doubling route to chaos. Even in these regimes, the stationary populations estimate the time-averaged flavor populations to within $10\%$ in our tests. These fixed points provide a physically motivated input for incorporating CFI into astrophysical simulations, with implications for neutrino heating and ejecta nucleosynthesis.
\end{abstract}

\maketitle

\textbf{\textit{Introduction.}}---Neutrino flavor conversion can change the exchange of energy and lepton number between neutrinos and matter, linking microscopic quantum dynamics to core-collapse supernovae (CCSNe) and binary neutron-star mergers (BNSMs). Global simulations using approximate flavor-conversion prescriptions show changes in neutrino heating and explosion dynamics in CCSNe~\cite{EhringEtAl2023Parametric,EhringEtAl2023Explosions,MoriEtAl2025ThreeD,WangBurrows2025Fast,AkahoEtAl2025Subgrid,WangBurrows2026Quasi,AkahoEtAl2026Bifurcated,GogilashviliTamborra2026Explodability,GogilashviliTamborra2026Failed}, as well as in remnant evolution, ejecta composition, and nucleosynthesis in BNSMs~\cite{LiSiegel2021,JustEtAl2022,FernandezEtAl2022,LundEtAl2025InSitu,QiuEtAl2025Merger,QiuEtAl2025Impact}. Changes in ejecta composition affect heavy-element production and kilonova emission, while changes in remnant dynamics can leave imprints on postmerger gravitational waves~\cite{LiSiegel2021,QiuEtAl2025Impact}. Uncertainty in the nonlinear flavor populations therefore propagates into predictions of astrophysical dynamics and multimessenger signals.

Extensive studies of neutrino radiation fields have investigated conditions for fast flavor instability (FFI) in CCSNe~\cite{TamborraEtAl2017Angular,AbbarEtAl2019Occurrence,DelfanAzariEtAl2019Linear,NagakuraEtAl2019Asymmetric,MorinagaEtAl2020Preshock,DelfanAzariEtAl2020Inside,AbbarEtAl2020Asymmetric,GlasEtAl2020Convection,AbbarEtAl2021Characteristics,CapozziEtAl2021Muon,NagakuraEtAl2021Where,HaradaNagakura2022Rotating,AbbarHaradaNagakura2025,ZaizenEtAl2024Cooling} and BNSM remnants and their accretion disks~\cite{WuTamborra2017,WuEtAl2017Nucleosynthesis,GeorgeEtAl2020HMNS,Richers2022Metrics,FrousteyEtAl2024MomentLSA,MukhopadhyayEtAl2024Crossings,FrousteyKnellerMcLaughlin2025Closure}. Collisional flavor instability (CFI) has also been identified in both environments~\cite{LiuEtAl2023Universality,AkahoEtAl2024CFI,LiuEtAl2024Muon,NagakuraEtAl2025BNSM,FrousteyEtAl2026Modes,UrquillaEtAl2026Disk,NgEtAl2026ThreeFlavor}. Local and global quantum-kinetic calculations have followed nonlinear flavor evolution in supernova-motivated settings~\cite{NagakuraZaizen2022,ShalgarTamborra2023Advection,NagakuraZaizen2023Scales,Nagakura2023Heating,XiongEtAl2023Transport,ShalgarTamborra2023Decoupling,NagakuraZaizen2023Postshock,ShalgarTamborra2024CFI,XiongEtAl2024Evolution,ShalgarTamborra2024QKE,CorneliusEtAl2024TwoD,ShalgarTamborra2025ThreeFlavor,ZaizenNagakura2026Gradient,RouxEtAl2026LESA}
and merger-remnant
environments~\cite{PadillaGayEtAl2021Merger,GrohsEtAl2023Merger,
Nagakura2023GlobalBNSM,GrohsEtAl2024TwoMoment,
RichersEtAl2024Asymptotic,UrquillaEtAl2026Disk,
NgEtAl2026ThreeFlavor}.

The disparity between microscopic flavor-conversion scales and hydrodynamic scales motivates subgrid models that replace unresolved flavor evolution with effective transport prescriptions. For fast flavor conversion (FFC), analytical approaches have yielded approximate asymptotic-state predictions~\cite{XiongQian2021Stationary,BhattacharyyaDasgupta2021Depolarization,ZaizenNagakura2023,ZaizenNagakura2023Conservation,FiorilloRaffelt2026Quasilinear,GoimilGarciaTamborra2026Quasisteady,BhattacharyyaDasgupta2022Fate}, and population prescriptions and subgrid schemes have been tested against quantum-kinetic calculations~\cite{XiongEtAl2023Asymptotic,RichersEtAl2024Asymptotic,
NagakuraJohnsZaizen2024BGK,XiongEtAl2025Subgrid,AbbarWuXiong2024PINN}. A comparable predictive description of CFI is less developed. CFI has already been incorporated into CCSN simulations using an assumed equipartition state~\cite{WangEtAl2025CFI}, whereas nonlinear studies find flavor swaps and energy-dependent conversion under different collision models~\cite{KatoNagakuraJohns2024,Zaizen2025}. Inhomogeneous calculations of competing fast and collisional instabilities further show that collisions can alter the asymptotic outcome~\cite{TakahashiEtAl2026Competition}. Predicting the balance between flavor exchange and collisional restoration is therefore an important open problem.

CFI arises when differences in neutrino and antineutrino collision rates amplify coherence through collective self-interactions~\cite{Johns2023,
JohnsXiong2022,LinDuan2023,Xiong2023,LiuZaizenYamada2023}. Analytical descriptions based on flavor pendula and averaged slow dynamics have provided insight into its nonlinear evolution~\cite{JohnsRodriguez2023,FiorilloPadillaGayRaffelt2024}. Most directly, Froustey derived approximate asymptotic populations for a homogeneous, isotropic, monochromatic relaxation model, identifying a balance between flavor conversion and collisional population restoration~\cite{Froustey2025}. This progress raises two linked questions: how can collision-balanced profiles be constructed for general energy spectra and energy-dependent collision rates, and what do they predict when the nonlinear evolution does not settle into a stationary state?

In this Letter, we construct general homogeneous collective-precession (HCP) solutions of two-flavor quantum kinetic equations (QKEs) that relax both populations and coherences. Their populations remain stationary while the coherences rotate together, making them fixed points in a rotating frame~\cite{DuanFullerQian2009Symmetries,XiongQian2021Stationary}. We focus here on isotropic models to isolate the balance between flavor exchange and collisional restoration. In this setting, two self-consistency equations determine the profiles for arbitrary discrete or continuous energy spectra, including energy-dependent collision rates. The construction is therefore energy resolved from the outset; monochromatic examples make the root-counting relation and nonlinear dynamics particularly transparent. An exact spectral embedding carries their stationary-branch structure and collective dynamics, including period-doubling bifurcations and chaos, into multienergy systems with energy-independent collision rates and compatible target spectra.

Weak linear growth need not imply small flavor conversion, and collisional relaxation need not produce stationary saturation. Nevertheless, the HCP populations estimate time-averaged flavor occupations to within $10\%$ across the sampled periodic and chaotic regimes, without fitting or changing the prescription between regimes. Selected spatially dependent calculations also suggest that HCP populations remain useful estimates of box-averaged flavor occupations despite pronounced spatial structure and local angular anisotropy. The construction supplies an energy-resolved population input for developing CFI subgrid models that consistently couple unresolved flavor dynamics to classical transport~\cite{Johns2025Subgrid}.

\textbf{\textit{Model and stationary-state construction.}}---We consider an energy-resolved gas of two neutrino flavors, $e$ and $x$, and develop the HCP construction explicitly in the isotropic limit. The diagonal entries of the occupation matrix $\rho_E$ are flavor populations, and the off-diagonal entries describe coherence; barred quantities refer to antineutrinos. Writing $\rho_E=(P_{0,E}I+\bm P_E\cdot\bm\sigma)/2$, $P_{0,E}$ and $P_{z,E}$ are the sum and difference of the occupations, while $P_{T,E}=2|\rho_{ex,E}|$ measures coherence. Keeping neutrino self-interactions and collisional relaxation, but neglecting vacuum and matter terms, the homogeneous QKEs~\cite{SiglRaffelt1993,Vlasenko2014} are
\begin{equation}
\begin{aligned}
\dot\rho_E&=-i[H,\rho_E]+\tfrac12\{\Gamma_E,\rho_E^{\textrm{(eq)}}-\rho_E\},\\
H&=\tfrac12\bm D\cdot\bm\sigma,\qquad
\bm D=\int(\bm P_E-\bar{\bm P}_E)\,d\mu_E,
\end{aligned}
\label{eq:energy-qke}
\end{equation}
with the same $H$ and barred collision quantities for antineutrinos. The commutator redistributes flavor coherently. Collisions restore populations toward the prescribed diagonal target $\rho_E^{\textrm{(eq)}}=(q_EI+p_E\sigma_z)/2$ and damp coherence, with collision-rate matrix $\Gamma_E=\operatorname{diag}(\kappa_{e,E},\kappa_{x,E})$. The collision rates are bounded away from zero and the target occupations are physical. Collisions act locally in energy; different energies communicate through $\bm D$. The positive spectral measure has finite total weight and includes discrete quadratures, $\int X_E\,d\mu_E=\sum_jw_jX_{E_j}$ with $w_j>0$, as well as continuous spectra. The self-interaction coefficient is absorbed into the time unit.

In an HCP state, the populations and coherence magnitudes are constant while all coherences rotate at frequency $\Omega$: $\rho_E(t)=U_{\Omega t}\rho_E^\star U_{\Omega t}^{\dagger}$, with $U_\theta=\exp(-i\theta\sigma_z/2)$ and time-independent $\rho_E^\star$, and analogously for antineutrinos. We solve each energy mode's stationary response to a trial collective field, then require the responses to reproduce that field.

Choose the rotating-frame transverse axis so that $\bm D_\perp=R\hat{\bm x}$, and define $\mathcal A=R^2>0$ and $\Lambda=\Omega-D_z$. With $s_E=(\kappa_{e,E}+\kappa_{x,E})/2$, $d_E=(\kappa_{e,E}-\kappa_{x,E})/2$, and $a_E=s_E-d_E^2/s_E>0$, the local stationary response is
\begin{equation}
\begin{aligned}
K_E(\mathcal A,\Lambda)&=a_E(s_E^2+\Lambda^2)+s_E\mathcal A,\\
P_{z,E}^\star&=\frac{a_E(s_E^2+\Lambda^2)p_E}{K_E},\\
P_{0,E}^\star&=q_E+\frac{d_E}{s_E}(p_E-P_{z,E}^\star).
\end{aligned}
\label{eq:stationary-populations}
\end{equation}
These expressions determine the individual occupations $\rho_{ee,E}^\star=(P_{0,E}^\star+P_{z,E}^\star)/2$ and $\rho_{xx,E}^\star=(P_{0,E}^\star-P_{z,E}^\star)/2$, with analogous barred relations. For $m=0,1$, define
\begin{equation}
I_m(\mathcal A,\Lambda)=\int\left[\frac{s_E^m a_Ep_E}{K_E}-\frac{\bar s_E^m\bar a_E\bar p_E}{\bar K_E}\right]d\mu_E.
\label{eq:stationary-integrals}
\end{equation}
The reconstructed transverse field is $D_x=-R\Lambda I_0$ and $D_y=-RI_1$. Requiring $(D_x,D_y)=(R,0)$ gives
\begin{equation}
1+\Lambda I_0(\mathcal A,\Lambda)=0,\qquad
I_1(\mathcal A,\Lambda)=0.
\label{eq:stationary-self-consistency}
\end{equation}
Every root with $\mathcal A>0$ determines one physical HCP profile, apart from an arbitrary common phase, and every nonzero isotropic HCP solution is obtained this way. The frequency follows from $\Omega=D_z+\Lambda$. Only two unknowns must be solved for, regardless of the energy resolution or energy dependence of the collision rates. In the corresponding nonresonant monochromatic limit, it recovers the approximate population solution of Ref.~\cite{Froustey2025}, without having to choose the unstable mode or impose asymptotic elimination of the approximate growth rate. Reconstruction, physicality, and this limiting correspondence are detailed in Supplemental Material (SM), Sec.~S1.

\begin{figure*}[t]
     \centering
     \includegraphics[width=\textwidth]{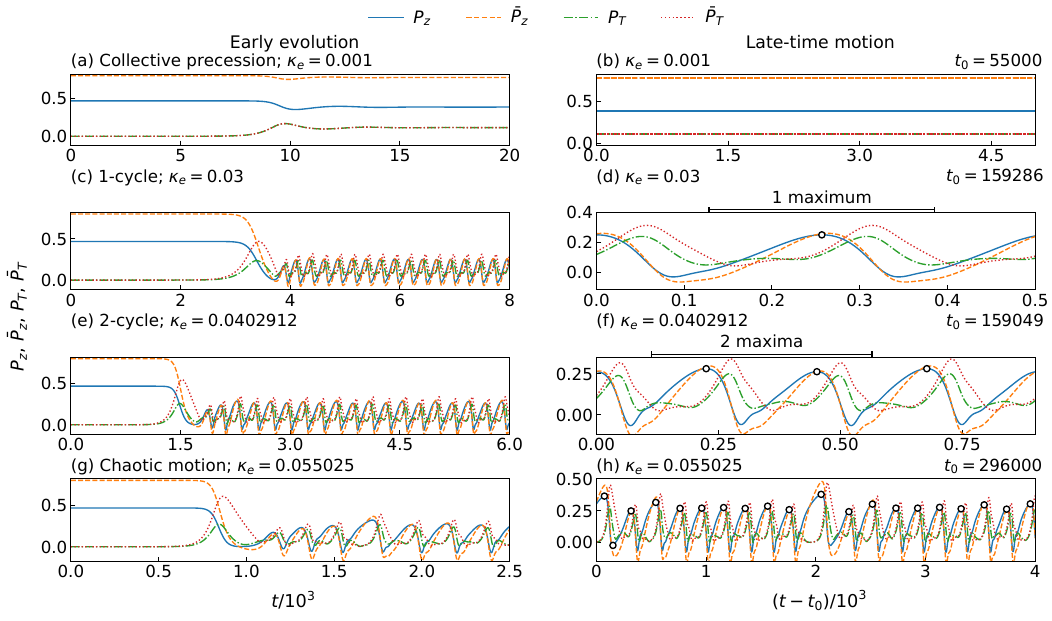}
     \caption{Flavor evolution from perturbed collision targets. The curves show population differences $P_z,\bar P_z$ and coherence magnitudes $\PT,\PbarT$. Each row pairs early and late segments of the same trajectory: (a,b) approach to HCP, (c,d) a 1-cycle, (e,f) a 2-cycle, and (g,h) chaotic motion. The brackets in (d,f) mark one period of the populations, coherence magnitudes, and relative phase, containing one and two $P_z$ maxima, respectively; the common flavor rotation is excluded. Open circles mark local $P_z$ maxima. The collision rate and late-time offset $t_0$ are indicated in each row.}
     \label{fig:evolution}
 \end{figure*}

Our monochromatic examples use one bin with unit weight and fix $(q,p)\simeq(0.834,0.469)$, $(\bar q,\bar p)\simeq(0.950,0.803)$, and $(\kappa_x,\bar\kappa_e,\bar\kappa_x)\simeq(7.62,13.5,3.92)\times10^{-3}$. Only the electron-neutrino collision rate $\kappa_e$ varies between runs; all parameters remain constant during each evolution. The trajectories in Fig.~\ref{fig:evolution} start at the collision-target populations with a coherence perturbation along the growing target mode, normalized to $[\PT^2(0)+\PbarT^2(0)]^{1/2}=10^{-8}$. Exact inputs and preparations are given in SM, Sec.~\ref{sm:numerics}.

\textbf{\textit{Monochromatic root count and stability.}}---For one energy bin with unit weight, the candidate states can be counted explicitly. Suppressing the energy label and writing $z=P_z^\star$, $\bar z=\bar P_z^\star$, and $S=s+\bar s$, the stationary equations become
\begin{equation}
\begin{aligned}
\bar z&=\bar p+\frac{a}{\bar a}(z-p),\\
\Pi(z)&=z\bar z S^2-s\bar s\bigl[S^2+(z+\bar z)^2\bigr]=0.
\end{aligned}
\label{eq:mono-quadratic}
\end{equation}
The first relation expresses collision balance between the two particle sectors and makes $\Pi$ at most quadratic in $z$. Each real root with positive reconstructed $\PT^2$ and $\PbarT^2$ gives one physical nonzero HCP solution, apart from its common phase.

The number of these solutions is linked to the target's linear instability. Population perturbations of the flavor-diagonal target decay, whereas coherence perturbations grow when $\mathcal C<0$, where
\begin{equation}
4\mathcal C=s\bar s\bigl[S^2+(p-\bar p)^2\bigr]-(s-\bar s)^2p\bar p.
\label{eq:target-CFI}
\end{equation}
The identity $\Pi(p)=-4\mathcal C$ connects the stationary polynomial to marginal CFI. Away from marginality and coincident roots, an unstable target has exactly one nonzero HCP solution, whereas a stable target has zero or two. The counting proof and stability analysis are given in SM, Sec.~\ref{sm:roots}. A root at the target has zero coherence and is excluded from this count. For arbitrary energy spectra, the zero-coherence boundary likewise reproduces the marginal CFI dispersion relation. The exact spectral embedding discussed below also carries the monochromatic count into its compatible multienergy family.

\begin{figure}[t]
    \centering
    \includegraphics[width=\columnwidth]{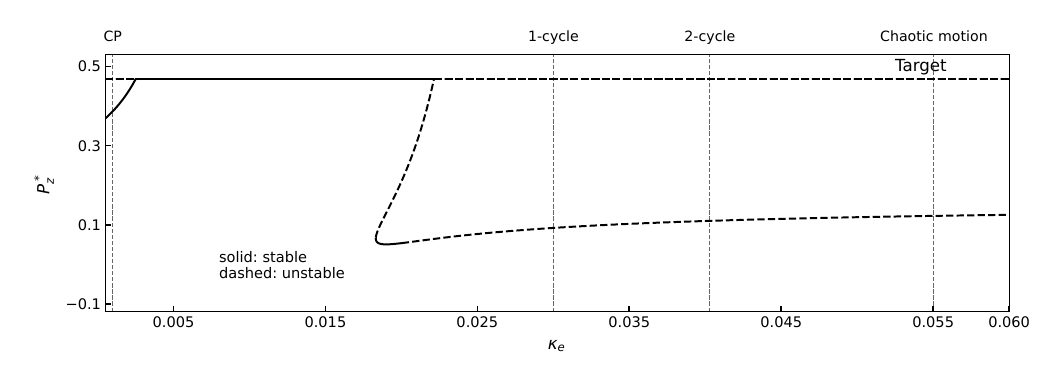}
    \caption{HCP branches and target stability versus $\kappa_e$. The horizontal line is the target $P_z=p$; other curves are HCP solutions. Solid and dashed segments denote attraction and instability, respectively, excluding the common phase for HCP. Labeled vertical guides mark the four examples of Fig.~\ref{fig:evolution}, not bifurcation thresholds.}
    \label{fig:equilibria}
\end{figure}

Existence does not imply attraction. We linearize the full QKEs in the rotating frame, including the response of the self-interaction field. Removing the common phase leaves a seven-dimensional system; negative real parts of all seven Jacobian eigenvalues imply local attraction. Figure~\ref{fig:equilibria} distinguishes attracting and unstable branches and locates the four trajectories of Fig.~\ref{fig:evolution}. The attracting HCP branches describe collision-balanced saturation of the kind predicted approximately in Ref.~\cite{Froustey2025}; coherence-damping-only models~\cite{Zaizen2025} have a different population balance.

The upper-left branch illustrates weak instability followed by weak conversion. As $\kappa_e$ approaches $0.0025$ from below, both the target growth rate and the population displacement of the attracting HCP solution vanish. A small perturbation thus saturates near the target. This continuous onset is a supercritical Hopf bifurcation.

At larger $\kappa_e$, a stable--unstable pair of finite-amplitude HCP solutions appears near $0.018$ while the target remains stable. The unstable member approaches the target and merges with it near $0.0221$, where the target destabilizes. This subcritical Hopf onset supplies no nearby attracting HCP state to limit conversion: a small positive growth rate can precede a finite-amplitude population change.

The other HCP branch has its own stability threshold. In this family it becomes unstable to oscillatory perturbations near $0.0200$, before the target destabilizes. These are distinct events involving different states; subcritical CFI onset does not itself require the surviving HCP solution to be unstable. Nor are the supercritical and subcritical onsets fixed by the near/far classification of the linear CFI mode: the collisional population response also matters.

\begin{figure}[t]
    \centering
    \includegraphics[width=\columnwidth]{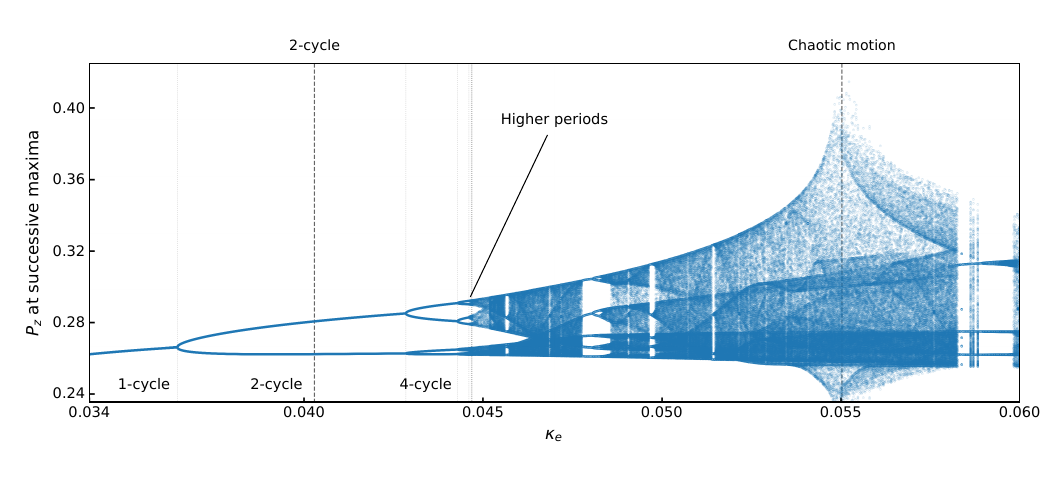}
    \caption{Period doubling and periodic windows in the monochromatic collision-rate scan. Points are successive local $P_z$ maxima from direct evolution and computed periodic orbits, including the 256-cycle. Dotted lines locate the eight resolved period doublings. The labeled vertical guides mark the 2-cycle and chaotic examples in Fig.~\ref{fig:evolution}; its HCP and 1-cycle examples lie below the displayed collision-rate interval. Only the principal positive-maximum bands are displayed.}
    \label{fig:branches}
\end{figure}

Coexistence makes the outcome preparation dependent. At $\kappa_e=0.0195$, a small target perturbation decays, whereas a finite-amplitude state inherited from the oscillatory regime approaches the attracting HCP solution. Across Fig.~\ref{fig:branches}, both the target and the unique nonzero HCP solution are unstable.

\begin{figure}[t]
    \centering
    \includegraphics[width=\columnwidth]{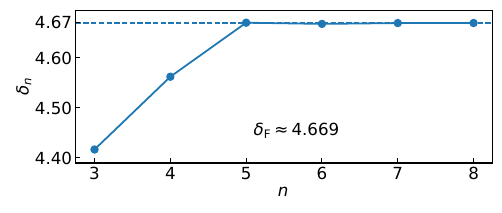}
    \caption{Feigenbaum-like spacing of the resolved period doublings. The ratios are $\delta_n=(\kappa_{n-1}-\kappa_{n-2})/(\kappa_n-\kappa_{n-1})$, where $\kappa_n$ is the $n$th period-doubling collision rate. The dashed line denotes the Feigenbaum constant, $\delta_{\rm F}\simeq4.669$.}
    \label{fig:cascade}
\end{figure}

\textbf{\textit{Period doubling and chaotic dynamics.}}---Figure~\ref{fig:branches} displays successive maxima of $P_z$ across $0.034\leq\kappa_e\leq0.060$. One recurring maximum splits into two, then four, and successively more before irregular motion develops; periodic windows reappear within the irregular regime. An $m$-cycle contains $m$ maxima before the populations, coherence magnitudes, and relative phase repeat, excluding the common flavor rotation. Periodic-orbit calculations resolve eight period doublings and a stable 256-cycle. Their spacing ratios approach the Feigenbaum constant $\delta_{\rm F}\simeq4.669$~\cite{Feigenbaum1978} (Fig.~\ref{fig:cascade}).

For the chaotic example at $\kappa_e=0.055025$, the leading Lyapunov estimate is $\widehat\lambda_1\simeq2.8\times10^{-3}$. Positive tangent growth persists under step refinement and throughout ten disjoint late-time blocks, establishing sustained sensitivity to nearby trajectories. Together with the bounded irregular population histories, this is strong numerical evidence for deterministic chaos. The periodic-orbit and Lyapunov checks are given in SM, Sec.~\ref{sm:numerics}.

\begin{figure}[t]
    \centering
    \includegraphics[width=\columnwidth]{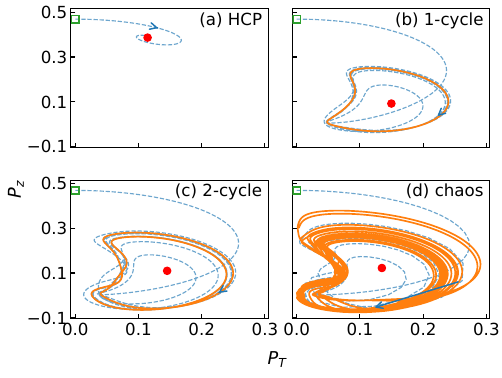}
    \caption{The trajectories of Fig.~\ref{fig:evolution} projected into $(P_T,P_z)$: (a) HCP, (b) 1-cycle, (c) 2-cycle, and (d) chaos. Dashed and solid curves show initial and late motion. Open squares mark the target; red points mark HCP solutions, attracting only in (a). These projections are not autonomous two-dimensional flows.}
    \label{fig:phase}
\end{figure}

Figure~\ref{fig:phase} shows relaxation to HCP in (a) and persistent excursions in (b)--(d). The 1-cycle extends toward the HCP oscillatory instability near $\kappa_e\simeq0.0200$, consistent with its origin as a modulation of that state. This motivates HCP as a reference for mean populations, whose accuracy we test directly.

For complete-period or bounded long-time averages, the mean occupations and HCP populations obey the same linear collision-balance constraints. Nonlinear correlations account for their differences (SM, Sec.~\ref{sm:averages}, which also specifies the averaging procedure). These relations motivate HCP as an estimator even without stationary saturation.

The independent HCP prediction differs from time-weighted trajectory means by at most $8.03\%$ for any species, attained by $\bar\nu_x$ at $\kappa_e=0.05515$ (SM, Fig.~\ref{fig:sm-population-estimator}). The same unique branch is used throughout, without fitting or changing the prescription between regimes. Thus HCP remains useful for mean-population prediction across the sampled periodic and chaotic regimes.

\textbf{\textit{Discussion and conclusions.}}---Our results distinguish the existence of collision-balanced states, their dynamical attraction, and their predictive value for mean populations. Weak linear CFI growth can produce finite conversion, while unstable HCP solutions can remain useful population estimators.

The construction is energy resolved from the outset. Moreover, the monochromatic branch structure and dynamics occur exactly in multienergy models with energy-independent collision-rate matrices and compatible spectra. Let $\varrho(t),\bar\varrho(t)$ be a monochromatic solution and choose
\begin{equation}
\begin{aligned}
\rho_E(t)&=c_E I+g_E\varrho(t),\\
\rho_E^{\textrm{(eq)}}&=c_E I+g_E\varrho^{\textrm{(eq)}},
\end{aligned}
\label{eq:spectral-embedding}
\end{equation}
with analogous barred expressions. The time-independent real envelopes preserve physical occupations, and $\int g_E\,d\mu_E=\int\bar g_E\,d\mu_E=1$ preserves the collective mean field. Substitution then gives the same collective evolution. Spectral-shape deviations with zero integral in each particle sector decay exponentially (SM, Sec.~\ref{sm:construction}). Hence the root count, collective stability, and bifurcations carry over to this family, while the additional spectral directions decay. General energy-dependent collision rates retain the HCP construction but can show richer dynamics.

HCP states are also exact solutions of spatially dependent QKEs with spatially uniform collision data. Selected one-dimensional periodic-box calculations approach the isotropic HCP solution with spontaneously generated spatial inhomogeneities rapidly decaying away, whereas others retain spatial structure and local angular anisotropy while their spatially averaged flavor occupations remain well approximated by the HCP populations with a maximum relative error around 10\%. This concerns diagonal populations, not equality of the averaged coherent state with HCP.

Using HCP populations in a subgrid model requires treating collisions consistently~\cite{NagakuraJohnsZaizen2024BGK,Johns2025Subgrid}. Adding an independent BGK relaxation toward the HCP profile to unchanged classical collisions risks counting population restoration twice: the HCP profile already incorporates its balance with flavor exchange. A predictor calculated with off-diagonal-only collisions~\cite{Zaizen2025} avoids incorporating that restoration in the auxiliary problem, but omits the replenishment that sustains the HCP balance. An alternative is to approximate the combined unresolved flavor and collision response. For the diagonal population matrix $F_E$, one candidate source is
\begin{equation}
\begin{aligned}
\mathcal S_E[F_E]&=\mathcal C_{{\rm cl},E}[F_E]\\
&\quad+\tfrac12\{\Gamma_{{\rm CFI},E},F_E^\star-F_E^{\textrm{(eq)}}\},
\end{aligned}
\label{eq:cfi-effective-source}
\end{equation}
where $F_E^\star=\operatorname{diag}(\rho_E^\star)$, $F_E^{\textrm{(eq)}}=\rho_E^{\textrm{(eq)}}$, and $\Gamma_{{\rm CFI},E}$ is an effective diagonal collision-rate matrix, set to zero in the absence of a modeled CFI correction. For the present relaxation model, $\mathcal C_{{\rm cl},E}[F_E]=\{\Gamma_E,F_E^{\textrm{(eq)}}-F_E\}/2$: choosing $\Gamma_{{\rm CFI},E}=\Gamma_E$ makes $F_E^\star$ its fixed point, while setting $\Gamma_{{\rm CFI},E}=0$ recovers classical transport. This illustrates how CFI can redirect collisional population restoration rather than add a second relaxation toward a collision-dependent state. The source in Eq.~\eqref{eq:cfi-effective-source} is one possible implementation. More generally, the HCP fixed points provide collision-balanced population estimates that should be useful across different CFI subgrid models.

The present contribution is the population prediction itself: a collision-balanced alternative to an imposed equipartition endpoint, applicable on arbitrary energy grids and useful beyond stationary saturation. Improving this input to transport addresses uncertainty in neutrino heating and ejecta composition, with consequences for explosion dynamics, nucleosynthesis, and kilonova emission~\cite{EhringEtAl2023Explosions,FujimotoNagakura2023Nucleosynthesis,LiSiegel2021,QiuEtAl2025Impact}.Stationary solutions can therefore remain physically useful even when the selected nonlinear state is not stationary.

\begin{acknowledgments}
\textbf{\textit{Acknowledgments.}}--- 
J.~L. is supported by Grant-in-Aid for JSPS Fellows (26KJ2052). 
L.~J. is supported by a Feynman Fellowship through LANL LDRD project number 20230788PRD1. 
H.~N. is supported by Grant-in-Aid for Scientific Research (23K03468), the NINS International Research Exchange Support Program, and the HPCI System Research Project (Project ID: 260058).
\end{acknowledgments}

\bibliographystyle{apsrev4-2}
\bibliography{refs}

\clearpage
\onecolumngrid
\phantomsection
\pdfbookmark[0]{Supplemental Material}{supplemental-material}
\begin{center}
\textbf{Supplemental Material}\par\vspace{0.4em}
\textbf{The Fate of Collisional Neutrino Flavor Conversion:}\\
\textbf{Fixed Points, Period Doubling, and Chaos}
\end{center}
\twocolumngrid
\setcounter{section}{0}
\setcounter{equation}{0}
\setcounter{table}{0}
\setcounter{figure}{0}
\setcounter{secnumdepth}{1}
\renewcommand{\thesection}{S\arabic{section}}
\renewcommand{\theequation}{S\arabic{equation}}
\renewcommand{\thetable}{S\Roman{table}}
\renewcommand{\thefigure}{S\arabic{figure}}
\renewcommand{\theHsection}{SM.\arabic{section}}
\renewcommand{\theHequation}{SM.\arabic{equation}}
\renewcommand{\theHtable}{SM.\arabic{table}}
\renewcommand{\theHfigure}{SM.\arabic{figure}}

\section{HCP construction and spectral extension}
\label{sm:construction}

We use the Letter's notation for homogeneous collective-precession (HCP) states. The spectral measure is finite and positive; targets and collision rates are bounded, with all four collision rates bounded below by a common $\kappa_*>0$. The occupation matrices are not normalized to unit trace. Physical states satisfy $0\preceq\rho_E\preceq I$, meaning that both eigenvalues lie in $[0,1]$; the prescribed diagonal targets satisfy the same bounds. Barred expressions follow by replacing neutrino quantities with their antineutrino counterparts.

\emph{Reconstruction and physicality.} In the rotating frame, $H_{\rm rot}=(R\sigma_x-\Lambda\sigma_z)/2$. The transverse stationary equations are $s_EP_{x,E}^\star=\Lambda P_{y,E}^\star$ and $s_EP_{y,E}^\star=-\Lambda P_{x,E}^\star-RP_{z,E}^\star$. Together with the diagonal populations in the Letter, they give
\begin{equation}
P_{x,E}^\star=-\frac{R\Lambda a_Ep_E}{K_E},\qquad
P_{y,E}^\star=-\frac{Rs_Ea_Ep_E}{K_E}.
\label{eq:sm-transverse}
\end{equation}
Integration yields $D_x=-R\Lambda I_0$ and $D_y=-RI_1$, hence the two self-consistency equations in the Letter. Positive collision rates imply that $R=0$ gives only the diagonal target; for $R>0$, the reconstruction is unique up to a common phase.

For a prescribed self-consistent field, define $B_E=-iH_{\rm rot}-\Gamma_E/2$. Since $B_E+B_E^\dagger=-\Gamma_E\preceq-\kappa_*I$, the unique local stationary matrix is
\begin{equation}
\rho_E^\star=\int_0^\infty e^{B_Et}\Gamma_E\rho_E^{\rm(eq)}
 e^{B_E^\dagger t}\,dt.
\label{eq:sm-physicality}
\end{equation}
The diagonal matrices $\Gamma_E$ and $\rho_E^{\rm(eq)}$ commute, so the integrand is positive semidefinite. Replacing the target by $I-\rho_E^{\rm(eq)}$ gives the same representation for $I-\rho_E^\star$. Thus every reconstructed root is physical. This fixed-field uniqueness does not imply uniqueness or attraction of the nonlinear self-consistency root.

\emph{Marginal instability.} Write transverse perturbations as $\delta P_{+,E}\propto e^{\lambda t}$, where $P_+=P_x+iP_y$, and let $D_T=\int(p_E-\bar p_E)\,d\mu_E$. Their collective dispersion function is
\begin{equation}
\begin{aligned}
\mathcal D(\lambda)=1+i\int\Biggl[
&\frac{p_E}{\lambda+s_E-iD_T}\\
-&\frac{\bar p_E}{\lambda+\bar s_E-iD_T}\Biggr]d\mu_E.
\end{aligned}
\label{eq:sm-dispersion}
\end{equation}
At $\lambda=i\Omega$ and $\Lambda=\Omega-D_T$, one obtains
$\mathcal D(i\Omega)=1+\Lambda I_0(0,\Lambda)+iI_1(0,\Lambda)$.
The zero-coherence boundary is therefore exactly the marginal CFI condition, including energy-dependent collision rates. This identity is not a global root-count theorem for arbitrary spectra.

\emph{Nonresonant correspondence.} For the monochromatic polynomial of the Letter,
\begin{equation}
\Pi(z)=-(sz-\bar s\bar z)(\bar s z-s\bar z)-s\bar s(s+\bar s)^2.
\label{eq:sm-factorization}
\end{equation}
In a regular weak-collision limit at fixed collision-rate ratios, with $s/\bar s\ne1$, nonzero limiting populations, and separated branches, the final term is subleading. The leading branch relations are $sz=\bar s\bar z$ or $\bar s z=s\bar z$. Combining these with $\tau=a(p-z)=\bar a(\bar p-\bar z)$ gives
\begin{equation}
\tau_-\simeq\frac{sp-\bar s\bar p}{s/a-\bar s/\bar a},\qquad
\tau_+\simeq\frac{\bar s p-s\bar p}{\bar s/a-s/\bar a},
\label{eq:sm-torque-limit}
\end{equation}
when the denominators are nonzero; only physical branches are retained. The occupation shifts from the collision targets obey
\begin{equation}
\begin{aligned}
\Delta f_e&=-\frac{\tau}{2\kappa_e},&
\Delta f_x&=\frac{\tau}{2\kappa_x},\\
\Delta\bar f_e&=-\frac{\tau}{2\bar\kappa_e},&
\Delta\bar f_x&=\frac{\tau}{2\bar\kappa_x}.
\end{aligned}
\label{eq:sm-shifts}
\end{equation}
Writing the number-density matrices of Ref.~\cite{Froustey2025} as $N=n_0\rho$ and $\bar N=n_0\bar\rho$, its $(\Gamma,\bar\Gamma,\mathfrak g,\bar{\mathfrak g})/\mu_0$ correspond to $(s,\bar s,z,\bar z)$, with $\mu_0=\sqrt2G_Fn_0$. For equal heavy-flavor collision rates and targets, Eqs.~\eqref{eq:sm-torque-limit}--\eqref{eq:sm-shifts} reproduce its population prediction. These relations follow from stationary balance without separately selecting a growing mode or eliminating an approximate growth rate.

\emph{Exact spectral embedding.} For energy-independent collision-rate matrices and the compatible affine targets in the Letter, take $\int g_E\,d\mu_E=\int\bar g_E\,d\mu_E=1$ and define
\begin{equation}
\varrho=\int(\rho_E-c_EI)\,d\mu_E,\qquad
r_E=\rho_E-c_EI-g_E\varrho,
\label{eq:sm-decomposition}
\end{equation}
with barred analogues. Then $\int r_E\,d\mu_E=\int\bar r_E\,d\mu_E=0$ separately. The collective matrices obey the monochromatic QKEs exactly, because the common Hamiltonian and collision matrices factor out of the energy integrals. The residuals satisfy
\begin{equation}
\begin{aligned}
\dot r_E&=-i[H(t),r_E]-\tfrac12\{\Gamma,r_E\},\\
\frac{d}{dt}\|r_E\|_{\rm F}^2
&=-2\operatorname{Tr}(\Gamma r_E^2)
\leq-2\kappa_*\|r_E\|_{\rm F}^2,
\end{aligned}
\label{eq:sm-residual-decay}
\end{equation}
where $\|r\|_{\rm F}^2=\operatorname{Tr}(r^\dagger r)$. Thus spectral-shape residuals contract at least as $e^{-\kappa_*t}$ in the corresponding energy-integrated norm, even along a chaotic collective trajectory. An HCP residual would instead have constant norm under common precession, and must therefore vanish. Within this compatible physical family, HCP roots correspond to the monochromatic roots; their count and collective stability carry over, with additional decaying spectral directions. The same collective periodic or chaotic trajectories are embedded exactly. Energy-dependent collision rates generally break this dynamical reduction, but not the stationary construction.

\section{Root count and stability}
\label{sm:roots}

\emph{Counting physical roots.} Let $z=P_z^\star$, $u=P_T^\star$, $\delta=\phi-\bar\phi$, and $\tau=u\bar u\sin\delta$. The stationary equations give
\begin{equation}
\begin{aligned}
\tau&=a(p-z)=\bar a(\bar p-\bar z),\\
u^2&=z\tau/s,\qquad \bar u^2=\bar z\tau/\bar s,\\
\cos\delta&=\frac{z+\bar z}{s+\bar s}\sin\delta.
\end{aligned}
\label{eq:sm-mono-reconstruction}
\end{equation}
For $p,\bar p>0$, positive squared amplitudes require $z,\bar z,\tau>0$, hence $0<\tau<\tau_{\max}=\min(ap,\bar a\bar p)$. On this interval put $Q(\tau)=\Pi(p-\tau/a)$, with $\bar z=\bar p-\tau/\bar a$. The Letter's identity gives $Q(0)=-4\mathcal C$, while $Q(\tau_{\max})=-s\bar s[(s+\bar s)^2+(z+\bar z)^2]<0$. Since $Q$ is at most quadratic, there is exactly one root for $\mathcal C<0$, and zero or two for $\mathcal C>0$, away from double roots. Each interior root reconstructs the amplitudes and relative phase in Eq.~\eqref{eq:sm-mono-reconstruction} and is physical by Sec.~\ref{sm:construction}. Simultaneous sign reversal covers $p,\bar p<0$. If $p\bar p\leq0$, the same-sign amplitude condition is impossible and $\mathcal C>0$. A root at $\tau=0$ has zero coherence and is excluded; an interior double root describes the merger of two finite-amplitude branches.

\emph{Dynamical stability.} At the target the transverse matrix is
\begin{equation}
M_T=\begin{pmatrix}-s-i\bar p&ip\\-i\bar p&-\bar s+ip\end{pmatrix}.
\label{eq:sm-target-matrix}
\end{equation}
Its characteristic polynomial is $\lambda^2+(S-iD_T)\lambda+s\bar s-i(sp-\bar s\bar p)$, where $S=s+\bar s$ and $D_T=p-\bar p$. It has one growing root exactly when $\mathcal C<0$; the four longitudinal eigenvalues are the negatives of the four collision rates.

For nonzero coherence, removing the common phase gives the state $\bm x=(z,\bar z,P_0,\bar P_0,u,\bar u,\delta)$ and the closed equations
\begin{equation}
\begin{aligned}
\dot z&=-u\bar u\sin\delta+s(p-z)+d(q-P_0),\\
\dot{\bar z}&=-u\bar u\sin\delta
+\bar s(\bar p-\bar z)+\bar d(\bar q-\bar P_0),\\
\dot P_0&=s(q-P_0)+d(p-z),\\
\dot{\bar P}_0&=\bar s(\bar q-\bar P_0)+\bar d(\bar p-\bar z),\\
\dot u&=z\bar u\sin\delta-su,\\
\dot{\bar u}&=\bar z u\sin\delta-\bar s\bar u,\\
\dot\delta&=\left(z\frac{\bar u}{u}+\bar z\frac{u}{\bar u}\right)\cos\delta-(z+\bar z).
\end{aligned}
\label{eq:sm-quotient}
\end{equation}
Writing the right-hand side as $\bm F_{\rm q}$, the real $7\times7$ Jacobian is $J_{\rm q}=\partial\bm F_{\rm q}/\partial\bm x$. At an HCP fixed point, negative real parts of all seven eigenvalues imply local attraction modulo the common phase; a positive real part implies instability. The equations include the perturbed mean field, not a fixed-field response. The common-phase zero direction is absent, but physical marginal eigenvalues are retained. The target, where this polar chart is singular, is tested with Eq.~\eqref{eq:sm-target-matrix} and the longitudinal block instead.

\emph{Two onsets and the separate HCP instability.} For a simple target Hopf point with $p>0$, normalize the critical amplitude by $P_+=w+O(|w|^3)$. Eliminating the quadratic population response gives
\begin{equation}
\begin{aligned}
\dot w&=(\gamma+i\Omega)w+c|w|^2w+\cdots,\\
\operatorname{Re}c&=-\frac{s\Pi'(p)}{apS[S^2+(2\Omega_c-D_T)^2]}.
\end{aligned}
\label{eq:sm-hopf}
\end{equation}
Here $\gamma=\operatorname{Re}\lambda$, $\Omega_c=(sp-\bar s\bar p)/S$, and $\Pi'(p)$ is taken along the collision-balance line. At the lower and upper target thresholds, $\operatorname{Re}c\simeq-0.046968$ and $+0.011022$, respectively; the other six target directions decay. The radial equation $\dot{|w|}=\gamma|w|+\operatorname{Re}c\,|w|^3+\cdots$ therefore gives the supercritical and subcritical onsets described in the Letter.

The near/far label, referring to the shifted frequency $\Omega-D_T$, does not fix this sign. Keeping $s,\bar s,p,\bar p$ unchanged preserves $M_T$, whereas changing $d$ or $\bar d$ changes $a$ or $\bar a$ and the nonlinear population response. With the other inputs fixed, equal barred collision rates $\bar\kappa_e=\bar\kappa_x=0.0086846765098$ make the upper far-mode onset supercritical ($\operatorname{Re}c\simeq-0.029583$). Choosing $(\bar\kappa_e,\bar\kappa_x)=(0.01650088536862,0.00086846765098)$ instead makes the lower near-mode onset subcritical ($\operatorname{Re}c\simeq0.025266$). These local comparisons do not alter the Letter's scan parameters.

The separate HCP crossing at $\kappa_e\simeq0.0200194578$ has critical quotient eigenvalues $\pm0.02037838\,i$, with the other five having negative real parts. It precedes target destabilization at $\kappa_e\simeq0.0221481171$ in this family. Neither its cubic criticality nor the exact cycle-to-HCP attachment is assigned. At $\kappa_e=0.0195$, the reported coexistence compares a fresh target perturbation with a finite-amplitude state carried downward from the oscillatory regime, rather than two infinitesimal target seeds.

\section{Numerical setup and evidence}
\label{sm:numerics}

\emph{Inputs and preparations.} With $t=\mu_0t_{\rm phys}$ and $\kappa_a=\Gamma_{a,\rm phys}/\mu_0$, the fixed monochromatic inputs are
\begin{align*}
(q,p)&=(0.8341403488,0.4690688184),\\
(\bar q,\bar p)&=(0.9503153116,0.8028357092),\\
\kappa_x&=0.007619307568,\\
\bar\kappa_e&=0.013453126276,\quad
\bar\kappa_x=0.0039162267436.
\end{align*}
Only $\kappa_e$ changes between evolutions. Target-seeded trajectories start with $(P_0,P_z,\bar P_0,\bar P_z)=(q,p,\bar q,\bar p)$ and $(P_+,\bar P_+)^{\mathsf T}=10^{-8}\bm v$, where $\bm v^\dagger\bm v=1$ and $\bm v$ is the leading right eigenvector of Eq.~\eqref{eq:sm-target-matrix}. The common seed phase is immaterial. Inherited finite-amplitude preparations and constructed periodic orbits are treated separately. The coexistence result is reproduced at Cartesian RK4 steps $0.025$ and $0.0125$.

The unified scan spans $0.034\leq\kappa_e\leq0.060$, with nonuniform sampling around the cascade. The uniform subset uses $\kappa_e=0.0452+i(2.5\times10^{-5})$, $i=0,\ldots,592$. At these 593 values, target-seeded RK4 integrations use steps $h=0.025$ and $0.0125$ to $t=300000$; the diagnostic interval is $200000<t\leq300000$. Cascade records retain their individual evolution durations and late windows, rather than being assigned this uniform-subset protocol.

\emph{Periodic orbits.} Multiple shooting solves $\Phi^T(\bm x_0)=\bm x_0$ in the quotient, fixing the time origin at a $P_z$ maximum. The monodromy matrix is obtained by integrating $\dot Y=J_{\rm q}(\bm x(t))Y$, $Y(0)=I_7$. Excluding the time-translation multiplier $+1$, attraction requires all Floquet multipliers inside the unit circle; a real multiplier crossing $-1$ identifies each doubling. Primary/tighter integrations use relative tolerances $10^{-11}/10^{-12}$, absolute tolerances $10^{-13}/10^{-14}$, and maximum steps $0.2/0.1$. Shooting residuals are below $5.57\times10^{-12}/1.74\times10^{-13}$. Both configurations resolve all eight crossings in Table~\ref{tab:sm-crossings}.

\begin{table}[t]
\caption{Period-doubling collision rates and spacing ratios. The parent cycle has $2^{n-1}$ maxima; $\kappa_n$ is the tighter-bracket midpoint. $W_n$ is the width of the interval containing both numerical crossing brackets; it is an empirical localization span, not a statistical error bar. Ratios use the unrounded coordinates.}
\label{tab:sm-crossings}
\centering
\setlength{\tabcolsep}{3pt}
\begin{tabular}{@{}rrrr@{}}
\toprule
$n$ & $\kappa_n$ & $W_n$ & $\delta_n$\\
\midrule
1 & $0.036460665251017$ & $9.54\!\times\!10^{-12}$ & ---\\
2 & $0.042844835712016$ & $7.15\!\times\!10^{-12}$ & ---\\
3 & $0.044290231805682$ & $8.58\!\times\!10^{-12}$ & $4.41690$\\
4 & $0.044607064957619$ & $3.81\!\times\!10^{-12}$ & $4.56201$\\
5 & $0.044674912561059$ & $1.19\!\times\!10^{-12}$ & $4.66978$\\
6 & $0.044689449216515$ & $5.36\!\times\!10^{-13}$ & $4.66735$\\
7 & $0.044692562633748$ & $7.75\!\times\!10^{-14}$ & $4.66904$\\
8 & $0.044693229441543$ & $1.55\!\times\!10^{-13}$ & $4.66914$\\
\bottomrule
\end{tabular}
\end{table}

Continuous full-period integrations and proper-divisor recurrence tests of the full phase-invariant state verify primitive periods, rather than repeated shorter orbits. At $\kappa_e\simeq0.044693234419$, the primitive 256-cycle has $T\simeq5.60648\times10^4$ and largest nonneutral Floquet modulus $0.93535$ in both configurations. Full-period closures are below $4.5\times10^{-13}$, while the smallest proper-divisor mismatch is about $1.72\times10^{-5}$. These tests provide strong numerical evidence for a locally attracting primitive 256-cycle. Target-seeded selection is independently confirmed at 25 tested parameters through 32-cycle motion, using longer integrations with the eighth-order DOP853 tableau at steps $0.5$ and $0.25$. Long-lived alternating transients near a flip are distinguished from doubled cycles by full-state recurrence and these extended integrations. Periodic-window assignments on the uniform grid require the same primitive count in both 50000-unit diagnostic halves and at both RK4 steps.

\emph{Lyapunov growth.} The full Cartesian state and tangent equations $\dot V=J_8(\bm y(t))V$ use identical RK4 stages. Tangent alignment occupies $180000\leq t\leq200000$, followed by the diagnostic interval above. Every $0.5$ time units, we factor $V=Q_kR_k$ and restart the tangent evolution with $Q_k$, giving~\cite{BenettinEtAl1980}
\begin{equation}
\widehat\lambda_j[t_a,t_b]
=\frac{\sum_{t_a<t_k\leq t_b}\log|R_{jj}^{(k)}|}{t_b-t_a}.
\label{eq:sm-lyapunov}
\end{equation}
The first QR column, leading over the full interval, is followed in every block. Table~\ref{tab:sm-lyapunov} shows positive leading growth throughout all ten disjoint 10000-unit blocks at both steps. At $\kappa_e=0.0452$, an independent adaptive DOP853 calculation in quotient coordinates gives $9.07\times10^{-4}$ and Cartesian RK4 with $h=0.0125$ gives $8.88\times10^{-4}$, each after discarding 20000 units and measuring for 60000 units. These separate controls corroborate positive nonlinear growth; they are not repeats of the $0.055025$ benchmark.

\begin{table}[t]
\caption{Lyapunov estimates at $\kappa_e=0.055025$ over $200000<t\leq300000$. The last two columns are the minimum and maximum over ten disjoint 10000-unit blocks. All estimates are multiplied by $10^3$.}
\label{tab:sm-lyapunov}
\centering
\begin{tabular}{@{}rrrr@{}}
\toprule
$h$ & Full interval & Minimum block & Maximum block\\
\midrule
$0.025$ & $2.97297$ & $2.07268$ & $3.32837$\\
$0.0125$ & $2.81675$ & $2.36786$ & $3.46248$\\
\bottomrule
\end{tabular}
\end{table}

\section{Time-averaged populations}
\label{sm:averages}

\begin{figure}[t]
    \centering
    \includegraphics[width=\columnwidth]{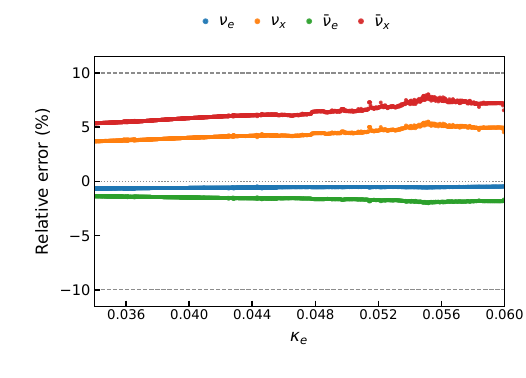}
    \caption{Relative errors of the HCP population estimate. For each species $\alpha\in\{\nu_e,\nu_x,\bar\nu_e,\bar\nu_x\}$, the plotted quantity is $100\epsilon_\alpha$, with $\epsilon_\alpha=(f_\alpha^\star-\langle f_\alpha\rangle_T)/\langle f_\alpha\rangle_T$; positive values denote overestimation. The time-weighted means come from direct-trajectory records over $0.034\leq\kappa_e\leq0.060$. The HCP prediction is calculated independently using the same branch throughout. Horizontal reference lines mark zero and $\pm10\%$. The maximum absolute discrepancy is $8.03\%$ for $\bar\nu_x$ at $\kappa_e=0.05515$.}
    \label{fig:sm-population-estimator}
\end{figure}

\emph{Comparison procedure.} Figure~\ref{fig:sm-population-estimator} uses 1594 direct-trajectory records: 1001 cascade histories and 593 half-step histories on the uniform grid. The four occupations are $f_e=(P_0+P_z)/2$, $f_x=(P_0-P_z)/2$, and their barred counterparts. For each record,
$\langle f\rangle_T=T^{-1}\int_{t_0}^{t_0+T}f(t)\,dt$
is evaluated with its actual timestamps, not by equally weighting maxima. Uniform-grid records use the late interval in Sec.~\ref{sm:numerics}, or the retained complete-period subinterval where applicable. Cascade histories use their retained late windows. Five averaging-sensitive cascade records at $\kappa_e=0.0446256$, $0.0446372$, $0.0446488$, $0.0446720$, and $0.0446836$ use the largest even number of complete recorded periods ending at the final saved maximum within the last half of their available continuation. Last-quarter and between-half comparisons test window sensitivity; the largest between-half sensitivity in these five checks is $0.1193\%$, and the largest quadrature sensitivity is $0.3112\%$.

Constructed-orbit means and primary-step controls are assessed separately, not added to the 1594 plotted records. Complete-period orbit averages use time-weighted integrals over their original shooting segments. The largest primary/tighter mean difference is $0.000977\%$ and selected-profile quadrature sensitivity is $0.01766\%$ for these orbit controls. The HCP occupations are computed independently from the same collision targets and collision rates. The quoted estimator error applies to these finite-record means, with averaging sensitivities assessed separately.

\emph{Averaged collision balance.} Define $Z=\bar P_xP_y-\bar P_yP_x=u\bar u\sin\delta$ and $\mathcal B_X=[X(t_0+T)-X(t_0)]/T$. Averaging the diagonal equations gives
\begin{equation}
\begin{aligned}
\langle f_e\rangle_T&=f_e^{\rm(eq)}-\frac{\langle Z\rangle_T}{2\kappa_e}
-\frac{\mathcal B_{f_e}}{\kappa_e},\\
\langle f_x\rangle_T&=f_x^{\rm(eq)}+\frac{\langle Z\rangle_T}{2\kappa_x}
-\frac{\mathcal B_{f_x}}{\kappa_x},
\end{aligned}
\label{eq:sm-mean-balance}
\end{equation}
with barred counterparts and the same $Z$. Boundary terms vanish for complete periods and tend to zero for bounded long-time records. HCP occupations obey the same relations with the constant torque $\tau^\star$ and no boundary term. Thus the mean and HCP occupations lie on the same collision-balance line in that limit. Their difference depends on the mean coherent torque, not on whether the trajectory becomes stationary.

Nonlinear correlations prevent replacing the full dynamics by equations for the means alone. For example,
\begin{equation}
\begin{aligned}
s\bigl[\langle u\rangle_T^2+\operatorname{Var}_T(u)\bigr]
&=\langle z\rangle_T\langle Z\rangle_T
+\operatorname{Cov}_T(z,Z)\\
&\quad-\tfrac12\mathcal B_{u^2},
\end{aligned}
\label{eq:sm-correlations}
\end{equation}
where the HCP relation is $s(u^\star)^2=z^\star\tau^\star$. These fluctuation terms explain why an HCP estimate need not equal the time average. Under the exact spectral embedding, $f_{\alpha,E}^\star-\langle f_{\alpha,E}\rangle_T=g_E(f_\alpha^\star-\langle f_\alpha\rangle_T)$, with barred analogues. Absolute discrepancies transform directly; relative discrepancies also depend on $c_E$ and need not retain the monochromatic percentage.

\end{document}